\pdfoutput=1
\documentclass[11pt,twoside,a4paper,cmspaper,final,collab]{cms-tdr}
\def\svnVersion{b1605b8}\def\svnDate{2022/11/17}\def\cmsCernNoTag{CERN-EP-2020-020}\def\cmsCernDate{\today}\def\cmsMessage{Published in Physics Letters B as \href{http://dx.doi.org/10.1016/j.physletb.2022.137566}{\doi{10.1016/j.physletb.2022.137566}.}}
\begin{document}\cmsNoteHeader{B2G-20-003}

\newcommand{\dbbone}{\ensuremath{D^{\PQb\PQb}_{\mathrm{j}_1}}\xspace}
\newcommand{\dbbtwo}{\ensuremath{D^{\PQb\PQb}_{\mathrm{j}_2}}\xspace}
\newcommand{\ptjtwo}{\ensuremath{\pt{}_{\mathrm{j}_2}}\xspace}
\newcommand{\mjtwo}{\ensuremath{m_{\mathrm{j}_2}}\xspace}
\newcommand{\masym}{\ensuremath{m_{\text{asym}}}\xspace}
\newcommand{\Mjj}{\ensuremath{M_{\mathrm{jj}}}\xspace}
\newcommand{\Deta}{\ensuremath{\abs{\Delta\eta}}\xspace}
\ifthenelse{\boolean{cms@external}}{\providecommand{\cmsLeft}{upper\xspace}}{\providecommand{\cmsLeft}{left\xspace}}
\ifthenelse{\boolean{cms@external}}{\providecommand{\cmsRight}{lower\xspace}}{\providecommand{\cmsRight}{right\xspace}}

\providecommand{\cmsTable}[1]{\resizebox{\linewidth}{!}{#1}}
\providecommand{\cmsSmallTable}[1]{\ifthenelse{\boolean{cms@external}}{\resizebox{\linewidth}{!}{#1}}{#1}}

\cmsNoteHeader{B2G-20-003}
\title{Search for new particles in an extended Higgs sector with four \texorpdfstring{\PQb}{b} quarks in the final state at \texorpdfstring{$\sqrt{s} = 13\TeV$}{sqrt(s) = 13 TeV}}

\date{\today}

\abstract{
A search for a massive resonance \PX decaying to a pair of spin-0 bosons \PGf that themselves decay to pairs of bottom quarks, is presented. The analysis is restricted to the mass ranges $m_\PGf$ from 25 to 100\GeV and $m_\PX$ from 1 to 3\TeV. For these mass ranges, the decay products of each \PGf boson are expected to merge into a single large-radius jet. Jet substructure and flavor identification techniques are used to identify these jets. The search is based on CERN LHC proton-proton collision data at $\sqrt{s}=13\TeV$, collected with the CMS detector in 2016--2018, corresponding to an integrated luminosity of 138\fbinv. Model-specific limits, where the two new particles arise from an extended Higgs sector, are set on the product of the production cross section and branching fraction for $\PX \to \PGf\PGf \to (\PQb\PAQb)(\PQb\PAQb)$ as a function of the resonances' masses, where both the $\PX \to \PGf\PGf$ and $\PGf \to \PQb\PAQb$ branching fractions are assumed to be 100\%. These limits are the first of their kind on this process, ranging between 30 and 1\unit{fb} at 95\% confidence level for the considered mass ranges.
}

\hypersetup{
pdfauthor={CMS Collaboration},
pdftitle={Search for new particles in an extended Higgs sector with four b quarks in the final state at sqrt(s) = 13 TeV},
pdfsubject={CMS},
pdfkeywords={CMS, Higgs, BSM, 2HDM}}

\maketitle

\section{Introduction}
The discovery, by both the ATLAS and CMS Collaborations~\cite{201230,chatrchyan_2013_observation,20121}, of a particle consistent with the Higgs boson (\PH) with standard model (SM) couplings, puts emphasis on searches for partners of this new boson, which are predicted in models with extended Higgs sectors. Such extensions, proposed in a variety of new physics models~\cite{th1,th2,th3,th4,th5,th6,th7}, generally postulate that an additional approximate global symmetry with spontaneous symmetry breaking is a sufficient condition for the existence of new spin-0 particles, \PX and \PGf, whose masses $m_\PX$ and $m_\PGf$ are not constrained. If $m_\PX > 2m_\PGf$, then $\PX \to \PGf\PGf$ is the dominant decay of the heavier particle, and \PGf couples to fermions similarly to \PH.

\begin{figure}[ht]
    \centering
    \includegraphics[width=0.48\textwidth]{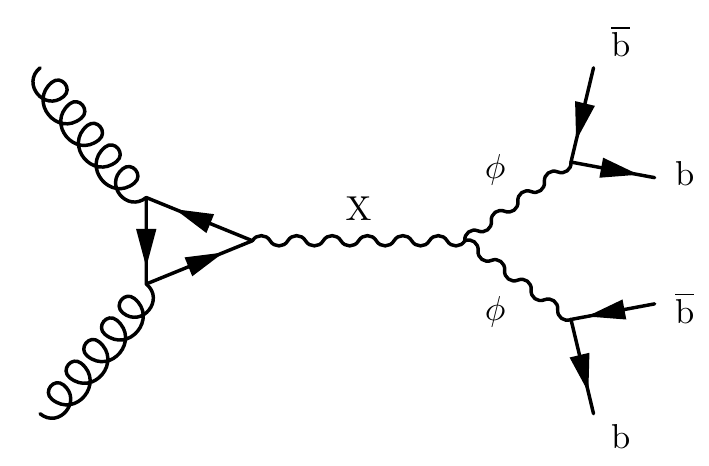}
    \caption{Feynman diagram of the production and decay of $\PX \to \PGf\PGf \to (\PQb\PAQb)(\PQb\PAQb)$.  The dominant production mechanism occurs via a fermion loop as shown in the diagram.  Additional partons may be present, produced by initial-state or final-state radiation.}
    \label{fig:Xaadiag}
\end{figure}

This paper presents a search for the process $\Pp\Pp \to \PX \to \PGf\PGf \to (\PQb\PAQb)(\PQb\PAQb)$ for $m_\PX$ between 1 and 3\TeV, and $m_\PGf$ between 25 and 100\GeV (for which the $\PGf \to \PQb\PAQb$ decay is expected to dominate). The leading Feynman diagram for this process is shown in Fig.~\ref{fig:Xaadiag}. In the considered model, the coupling of the boson \PX to gluons is evaluated by integrating over $N$ flavors of quarks that receive all their mass from the \PX vacuum expectation value $f$, such that the cross section depends only on the quantity $m_\PX N/f$. The search is based on proton-proton ($\Pp\Pp$) collision data at $\sqrt{s} = 13\TeV$ collected with the CMS detector~\cite{CMS:2008xjf} at the LHC in 2016--2018, corresponding to a total integrated luminosity of 138\fbinv~\cite{lumi2016,lumi2017,lumi2018}. In the mass ranges considered for the \PX and \PGf scalar bosons, the high momentum imparted to the \PGf boson causes the hadronic showers of the two \PQb quarks to overlap, such that the signal is best reconstructed as a pair of large-radius jets each with substructure consistent with the decay to two \PQb quarks. The ATLAS and CMS Collaborations have previously performed searches~\cite{old1,old2,old3,old4,old5} for the process $\PX \to \PH\PH \to (\PQb\PAQb)(\PQb\PAQb)$ with similar topologies. However, requirements on the jet mass in those analyses make them less sensitive to $m_\PGf$ below 100\GeV.

This analysis searches for a localized excess in the two-dimensional distributions of average jet mass and dijet mass for events with two large-radius jets with heavy-flavor jet substructure. The main background, from SM events composed uniquely of jets produced through the strong interaction, referred to as quantum chromodynamics (QCD) multijet events, is derived using control samples in data. Additional background from  top quark pair (\ttbar) events is estimated from simulation, with corrections obtained from control regions in data. Other SM backgrounds, namely single top quark, single vector boson, and paired vector boson processes, were found to be negligible. Tabulated results are provided in the HEPData record for this analysis~\cite{hepdata}.

\section{The CMS detector}
The CMS apparatus~\cite{CMS:2008xjf} is a multipurpose, nearly hermetic detector, designed to trigger on~\cite{Khachatryan:2016bia} and identify electrons, muons, photons, and (charged and neutral) hadrons~\cite{CMS:2015xaf,CMS:2018rym,CMS:2015myp,CMS:2014pgm}. A global reconstruction ``particle-flow" (PF) algorithm~\cite{Sirunyan:2017ulk} combines the information provided by the all-silicon inner tracker and by the lead-tungstate crystal electromagnetic (ECAL) and brass-scintillator hadron calorimeters (HCAL), operating inside a 3.8\unit{T} superconducting solenoid, with data from gas-ionization muon detectors interleaved with the solenoid return yoke, to build \PGt leptons, jets, missing transverse momentum, and other physics objects~\cite{CMS:2018jrd,CMS:2016lmd,CMS:2019ctu}. Events of interest are selected using a two-tiered trigger system~\cite{Khachatryan:2016bia}. The first level, composed of custom hardware processors, uses information from the calorimeters and muon detectors to select events at a rate of around 100\unit{kHz} within a time interval of less than 4\mus. The second level, known as the high-level trigger, consists of a farm of processors running a version of the full event reconstruction software optimized for fast (online) processing, and reduces the event rate to around 1\unit{kHz} for use in (offline) analyses. A more detailed description of the CMS detector, including its coordinate system, can be found in Ref.~\cite{CMS:2008xjf}.

\section{Monte Carlo simulation}
The benchmark signal model $\PX \to \PGf\PGf \to (\PQb\PAQb)(\PQb\PAQb)$, where \PX is produced through gluon fusion, was generated to leading order with the \MADGRAPH 2.6.0~\cite{madgraph} generator. Both the $\PX \to \PGf\PGf$ and $\PGf \to \PQb \PAQb$ branching fractions are set to 100\%. The samples are generated with up to two additional initial-state or final-state radiation jets. 
The total production cross section is calculated numerically at next-to-next-to-leading order (NNLO) in the infinite fermion mass limit using the \textsc{HqT} 2.0 program~\cite{theory1,theory2,theory3}.

The \MADGRAPH generator is used to model at leading order the QCD background, which was used to optimize the analysis procedure. The \POWHEG 2.0~\cite{POW1,POW2,POW3} generator is used to model \ttbar events at next-to-leading order~\cite{Alioli:2011as}.

The parton showering and fragmentation for all signal and background samples is done with \PYTHIA 8~\cite{pythia8}, and matching between the matrix element and parton shower jets relies on the MLM matching procedure~\cite{mlmmatch}. The CUETP8M1 (CP5)~\cite{pythiaTune,m_2020_extraction} underlying event tune is used for the simulation corresponding to the 2016 (2017--2018) data taking with the  NNPDF3.0 (NNPDF3.1)~\cite{nnpdf} NNLO parton distribution function (PDF) sets.  

The simulation of the CMS detector for all samples is handled by \GEANTfour~\cite{geant}. All samples include effects of additional $\Pp\Pp$ interactions in the same or adjacent bunch crossings, referred to as pileup. The pileup distribution in simulation is weighted to match the one observed in data. To account for any small differences noted above, systematic uncertainties associated with the simulations are treated as uncorrelated between years.

\section{Event reconstruction}
Jets used in this analysis are clustered using \textsc{FastJet}~\cite{ref:fastjet} with the anti-\kt (AK) algorithm~\cite{Cacciari:2008gp} and a distance parameter of 0.4 (AK4 jets) or 0.8 (AK8 jets). Particles produced in additional collisions within the same bunch crossing are suppressed by applying a weight to each PF candidate, calculated by the pileup-per-particle identification~\cite{ref:puppi,ref:pu} algorithm. Jets are corrected as a function of their \pt and pseudorapidity ($\eta$) to match the observed detector response~\cite{jec}.

Jets arising from the hadronization of a \PQb quark-antiquark pair are identified using the ``double-\PQb tagger''algorithm~\cite{Sirunyan:2017ezt}, which uses a boosted decision tree (BDT) of several vertex- and track-related variables to identify jets containing two displaced vertices consistent with decays into $\PQb\PAQb$. The BDT was trained on samples of simulated of boosted Higgs boson jets, where the jet showering may be different from the behavior of jets in real data. To correct for this, a scale factor is derived and applied to each jet in the signal simulation as a function of the jet \pt~\cite{Sirunyan:2017ezt}. This measurement is performed using a sample of high-\pt jets enriched with gluon decays to \PQb quark-antiquark pairs. To account for different data-taking conditions, this training is performed independently for each year. As a result, the exact performance of the discriminator differs from year to year. The soft drop mass algorithm (with angular exponent $\beta=0$ and soft threshold $z_{\text{cut}}=0.1$)~\cite{Dasgupta:2013ihk,Larkoski:2014wba} is used to remove soft and wide-angle radiation from jets before computing a ``groomed jet mass'' ($m_\mathrm{j}$), which better reflects the mass of the particle that initiated the jet. The lower bound on \PGf masses considered is due to uncertainty in the behavior of the double-\PQb tagger for groomed jet masses below 25\GeV.

\section{Event selection}
Events are first selected by a trigger based on either the \pt of a single AK8 jet, or on the event \HT, defined as the scalar \pt sum of all AK4 jets with $\pt > 30\GeV$ and $\abs{\eta} < 2.5$. In 2016 the \HT threshold was set to 800\GeV for the early data-taking period and raised to 900\GeV for the last 8\fbinv. This threshold was raised to 1050\GeV for data taken in 2017 and 2018. The AK8 jet threshold was set to 450 (500)\GeV in 2016 (2017--2018). The efficiency of this trigger combination is measured in an orthogonal sample triggered on single-muon candidate events, and ranges from 42\% for samples with $m_\PX$ of 1\TeV to 100\% for $m_\PX$ above 1.2\TeV. To account for differences between the simulated and real responses of the detector, the ratio of the efficiency of the trigger between data and simulation is applied to all simulated events, with the uncertainty in the efficiencies considered as a systematic uncertainty in the shape and normalization of the signal.

Events are only considered if they pass the trigger for their given year, and if they contain two AK8 jets with $\pt > 300\GeV$ and $\abs{\eta}<2.4$. An offline selection of $\HT > 900\GeV$ is applied. Signals with $m_\PX$ below 1\TeV do not produce events with \HT large enough to be considered in this analysis. The jets are ordered according to their \pt.

Events are divided between several search and control regions based on the mass asymmetry between the two leading jets $\mathrm{j}_1$ and $\mathrm{j}_2$ ($\masym = \abs{m_{\mathrm{j}_1} - m_{\mathrm{j}_2}}/(m_{\mathrm{j}_1} + m_{\mathrm{j}_2})$), the separation in $\eta$ between these jets (\Deta), and the values of the double-\PQb tagger discriminant ($D^{\PQb\PQb}$) for each jet. Two signal enriched regions are used to extract the signal: the first, called the ``tight search region'' (TSR), requires $\Deta < 1.5$, $\masym < 0.1$, $\dbbone > 0.8$, and $\dbbtwo > 0.6$. The second, called the ``loose search region'' (LSR), differs only in the constraint on the mass asymmetry, requiring $\masym \in [0.1,0.25]$. This second region, while it has lower sensitivity to the signal, contains a large contribution from the \ttbar background, allowing for strong constraints to be placed on that process, thus benefiting both search regions. These selections, including the two threshold values for $D^{\PQb\PQb}$, were optimized by maximizing the signal significance~\cite{Punzi} with respect to simulated backgrounds. Two control regions are defined with respect to each search region: the ``$\Deta$ sideband'', where the $\Deta$ requirement is replaced by $\Deta>1.5$, and the ``double-\PQb sideband'', where the double-\PQb discriminant on the leading jet becomes $-0.8 < \dbbone < 0.3$. Both control regions are used to estimate the background in the search regions, as they are rich in background events, have similar kinematic distributions to the search regions, and are signal depleted. The categorization is summarized in Table~\ref{tab:cat}. The background estimate, described in the next section, uses events from an additional category: the ``failing region'', defined for each search or control region, consisting of events fulfilling all selection criteria except that the subleading jet has $\dbbtwo < 0.6$.

After all selection criteria are applied, the product of the acceptance and efficiency for the range of $m_\PX$ and $m_\PGf$ considered is between 5 and 20\%. The efficiencies are lowest for signals with $m_\PX=1.0\TeV$, peak at $m_\PX=1.5\TeV$, and fall linearly to approximately 12\% for $m_\PX=3.0\TeV$. The signal efficiencies have negligible dependence on $m_\PGf$.

\begin{table}[ht!]
\centering
\topcaption{Search and control regions used in the analysis. A selection on the subleading jet double-\PQb-tagger discriminant $\dbbtwo > 0.6$ further separates each region into ``passing'' and ``failing'' categories.}
\renewcommand{\arraystretch}{1.25}
\cmsSmallTable{\begin{tabular}{l c c c}
& $\masym$ & $\Deta$ & \dbbone \\\hline
Tight search region & $<$0.1 & $<$1.5 & $>$0.8 \\
Loose search region & $ [0.1, 0.25]$ & $<$1.5 & $>$0.8 \\
Tight $\Deta$ sideband & $<$0.1 & $>$1.5 & $>$0.8 \\
Loose $\Deta$ sideband  & $[0.1, 0.25]$ & $>$1.5 & $>$0.8 \\
Tight double-\PQb sideband & $<$0.1 & $<$1.5 & $[-0.8, 0.3]$ \\
Loose double-\PQb sideband  & $[0.1, 0.25]$ & $<$1.5 & $[-0.8, 0.3]$ \\
\end{tabular}}
\label{tab:cat}
\end{table}

\section{Background processes}
The QCD and \ttbar backgrounds are simultaneously evaluated in a maximum likelihood fit of the TSR and LSR in all years. In this section we describe the background estimation. In the following section we detail the maximum likelihood fit itself and the systematic uncertainties associated with it.

The \ttbar contribution to the total background is obtained from simulation, after a correction is applied to account for the differences~\cite{ttMC1, ttMC2, ttMC3} between the top quark pair transverse momentum distribution in fixed-order simulations and data. This variable is highly correlated with the dijet mass of the \ttbar system. The correction is applied by weighting the simulated events with a term: $W = \exp[\alpha (\pt^{\PQt} + \pt^{\PAQt})]$, where $\alpha$, initialized at $5\ten{-4}/\GeVns$, controls the shape of the resulting \ttbar distribution. The normalization of the \ttbar background and the value of $\alpha$ are constrained in the final simultaneous maximum likelihood fit of the TSR and LSR. The best fit values for both the overall normalization and the value of $\alpha$ may vary from year to year, so both are allowed to vary within large uncertainties (20 and 100\%, respectively), separately for each year. The uncertainty in the \ttbar normalization covers that associated with the application of the double-\PQb tagger to jets from that process, which may contain at most one genuine \PQb quark.

\begin{figure}[tbh!]
    \centering
        \includegraphics[width=\linewidth]{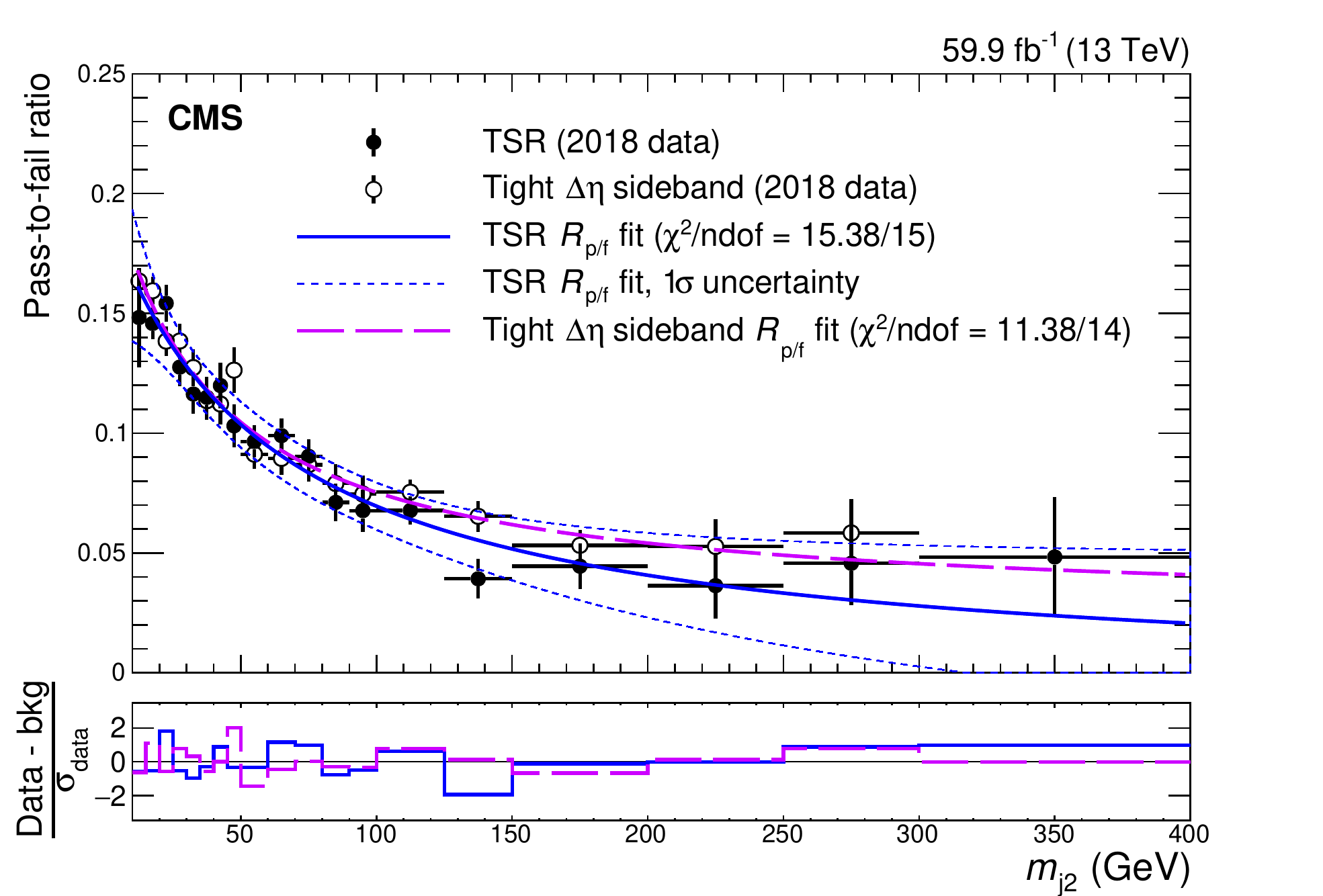}
    \caption{Distributions of the ratio of event passing and failing the $\dbbtwo > 0.6$ requirement, as a function of the subleading jet groomed mass $\mjtwo$, in 2018 data, for the TSR (filled markers) and the tight \Deta sidebands (open markers). The arctangent $R_{\text{p/f}}$ fit from which the background is estimated is shown by the solid line in the case of the TSR, and by the thick dashed line for the sideband, with the resulting $\chi^2$ and degrees of freedom indicated in the legend. The 1$\sigma$ uncertainty band of the fit in the TSR, from which systematic uncertainties on the QCD background estimate are derived, is shown by the thin dashed lines. Similar results were obtained for 2016 and 2017 data and in the three LSRs.  The lower panel shows the difference between the observed data and the fits, divided by the statistical uncertainty of the ratio of passing and failing events in the data for each bin.}
    \label{fits}
\end{figure}

The dominant QCD multijet background is modeled by exploiting the fact that the ratio of events for which the subleading jet passes or fails the $\dbbtwo > 0.6$ requirement in each search region can be modeled by a smooth function of the subleading jet \pt and groomed jet mass, $R_{\text{p/f}}(\ptjtwo,\mjtwo)$. This pass-to-fail ratio is parameterized as $R_{\text{p/f}} = P_n(\ptjtwo) F(\mjtwo)$, where $P_n$ is a polynomial of order $n$ and $F(\mjtwo)$ is the function $F(x) = p_0\arctan(p_1x + p_2) + p_3$. The parameters of each component of $R_{\text{p/f}}$ are initialized from their values in the control regions. The \Deta sidebands are used for $F(\mjtwo)$, whereas the double-\PQb sidebands are used for $P_n(\ptjtwo)$. In both cases the signal is negligible in the control regions. Figure~\ref{fits} shows, in 2018 data, the groomed mass dependence of $R_{\text{p/f}}$ for both the TSR and the $\Delta\eta$ sidebands, as well as the fit from which the background estimate is extracted; similar results were obtained for the other years and in the LSR. The QCD background in the ``passing'' search regions is estimated by weighting the events in the ``failing'' search regions by this $R_{\text{p/f}}$, after subtracting the expected contribution from \ttbar production. Systematic uncertainties in the \ttbar estimate mentioned above are propagated through to the QCD background estimate. This background estimate is also performed in each control region as a closure test, the results of which indicates that no additional systematic uncertainties need to be applied to the QCD background estimate.

To account for differences in the implementation of the double-\PQb tagger, each year of data taking, as well as the TSR and LSR, are treated independently. The order of $P_n$ is determined in the double-\PQb control region by performing Fisher F-tests~\cite{Ftest} on progressively higher-order polynomials. A $P_2$ ($P_1$) is found to be optimal for describing the data taken in 2016 (2017--2018). A third-degree polynomial (with the order chosen by a similar F-test) was also considered for modeling $F(\mjtwo)$: as it did not lead to a significantly improved description of $R_{\text{p/f}}$, it was reserved for bias tests of the background estimate (described below).  Uncertainties in the fit parameters for $F$ and $P_n$ in the signal regions are treated as systematic uncertainties in the shape and normalization of the QCD background in the search regions.

\section{Fitting procedure and systematic uncertainties}

A two-dimensional mass spectrum is examined for localized excesses of events, since both $m_\PGf$ and $m_\PX$ for a potential signal are unknown. The two dimensions of this spectrum are the average jet mass $\hat{m} = (m_{\mathrm{j}_1} + m_{\mathrm{j}_2})/2$ and the dijet mass \Mjj (the invariant mass of the sum of the two jet four-momenta). The signal is modeled analytically using a multivariate normal distribution whose five parameters (the mean and width of the distribution in $\hat{m}$ and \Mjj, and the correlation between the two masses) are fit to the values observed in the generated signal samples. The signal is sharply peaked in both variables, with root-mean-square values of 12--16\% and 6--9\% of the resonance mass for the reconstructed \PGf and \PX masses, respectively. The difference between generated signal shapes and those obtained from this functional form are negligible. The range of average jet masses considered is $\hat{m} \in [15, 200]\GeV$, while \Mjj between 0.9 and 5.0\TeV are considered. The $\hat{m}$ range extends beyond the signal masses considered in this analysis, with the events at high masses providing constraints on the \ttbar and QCD background components. Both mass distributions use variable binning to ensure that the background estimate prediction is nonzero in each mass interval.

Systematic uncertainties are treated in the fit as nuisance parameters affecting the shapes and the normalization of signal and background processes. All uncertainties are quantified in Table~\ref{tab:unc}. The dominant uncertainty is from the double-\PQb tagger scale factor, which is described in Ref.~\cite{Sirunyan:2017ezt} and applied as a function of the jet \pt. To account for differences between simulation and data on the jet energy calibrations, the jet energy scale (JES) and resolution uncertainties are considered as uncertainties in the shape of the signal and \ttbar background. The JES uncertainties take into account correlated and uncorrelated components between all three years. Uncertainty in the jet mass resolution was found to have a negligible effect on the result. Uncertainties pertaining to the trigger efficiency, the pileup distribution, the PDFs, and the integrated luminosity determination~\cite{lumi2016,lumi2017,lumi2018} are applied to the signal and \ttbar process simulations. The \ttbar background is additionally affected by the uncertainties in the normalization and the $\alpha$ parameter. The data-driven QCD background has two sources of uncertainty, from the determination of $R_{\text{p/f}}$, and from the statistical uncertainty in the failing regions. The latter dominates for high dijet and average jet masses where the number of events in the failing region is low.

\begin{table}[ht!]
\centering
\topcaption{Sources of systematic uncertainties considered in the analysis. The uncertainty in the integrated luminosity only affects the normalization; for the rest, both the shape and normalization are affected. The parameters affecting only the normalization have log-normal priors, and those affecting the shape (or both the shape and normalization) have Gaussian priors, except for the statistical uncertainty in the failing region, whose parameters were sampled from a $\Gamma$~distribution. Uncertainties marked with R are correlated between the TSR and LSR for a given year of data-taking, and those marked with Y are correlated between both search regions in all three years. All other uncertainties are uncorrelated between search regions. The values indicated in the table represent the pre-fit values of the uncertainty in the parameter. When a range is given, it indicates the typical variation of the size of the uncertainty over the average jet mass and dijet mass distribution. We note that all \ttbar uncertainties are propagated into the QCD backgroound estimate.}
\cmsSmallTable{\begin{tabular}{ l c c c c}
    & \multirow{2}{*}{Signal} & \multicolumn{2}{c}{Background} & \multirow{2}{*}{Corr.} \\
    & & \ttbar & QCD & \\ \hline  
    Integrated luminosity &    1.2--2.5\%    &          1.2--2.5\%             &             &  R  \\
    Double-$\PQb$ scale factor &    19--46\%     &                       &           &  R   \\
    Trigger efficiency &    1--5\%     &          1--5\%             &          &   R   \\
    PDF &    4\%    &          4\%             &        &     R   \\
    Pileup &    1--10\%    &          1--10\%             &         &   R    \\
    Jet energy scale (correlated) &    2\%    &            2\%           &         &    RY   \\
    Jet energy scale (uncorrelated) &   2\%     &         2\%              &          &   R   \\
    Jet energy resolution &    10\%    &          10\%             &           &  R   \\
    \ttbar shape ($\alpha$) &        &          100\%          &    100\%     &    R   \\
    \ttbar normalization  &        &           20\%            &     20\%     &  R    \\
    $R_{\text{p/f}}$ fit &        &                       &         5--30\%     & \NA \\
    Statistical uncertainty (failing region) &        &                       &       $<$1--100\%    &  \NA\\
\end{tabular}}
\label{tab:unc}
\end{table}

\begin{figure*}[tbh!]
    \centering
        \includegraphics[width=0.48\textwidth]{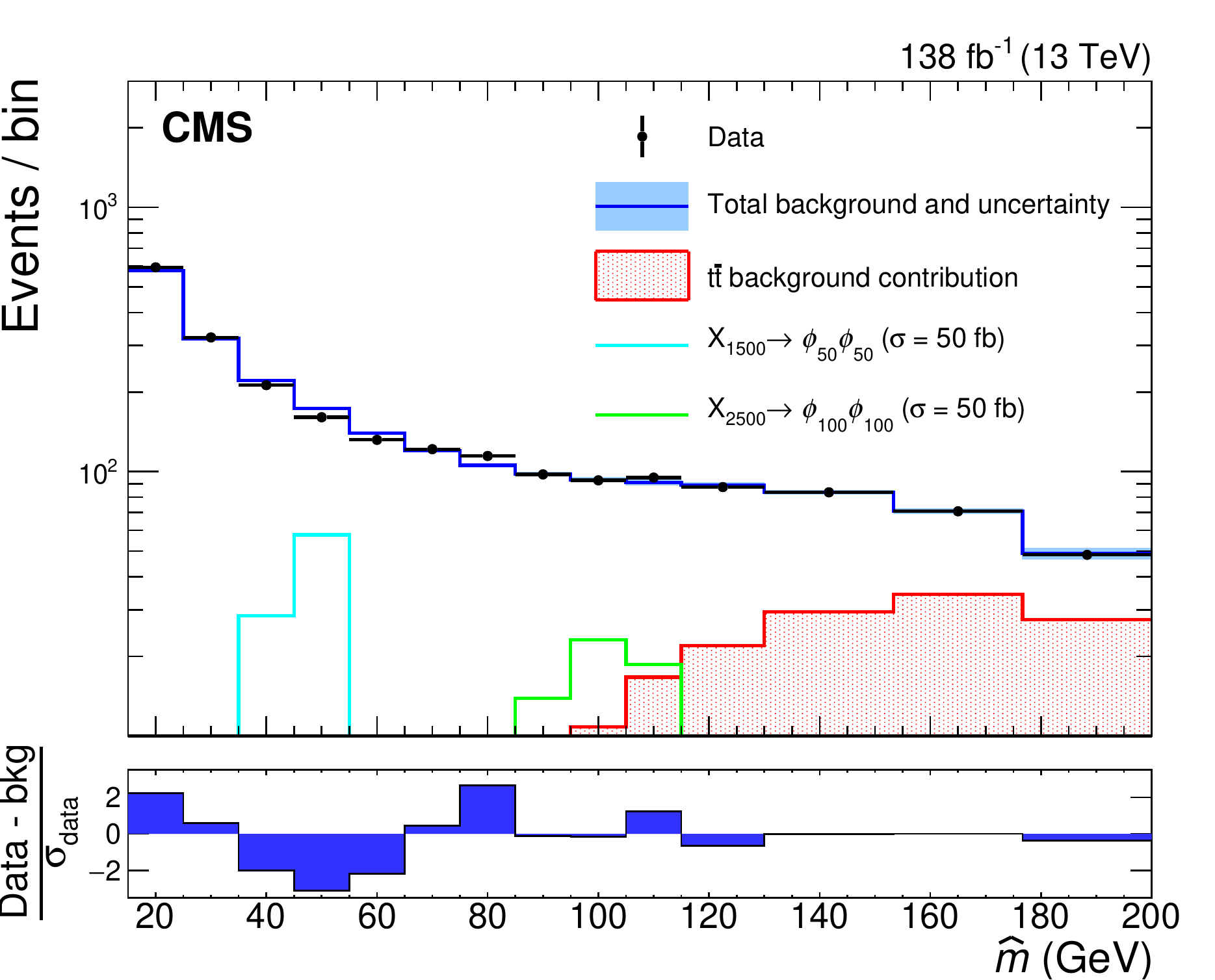}
        \includegraphics[width=0.48\textwidth]{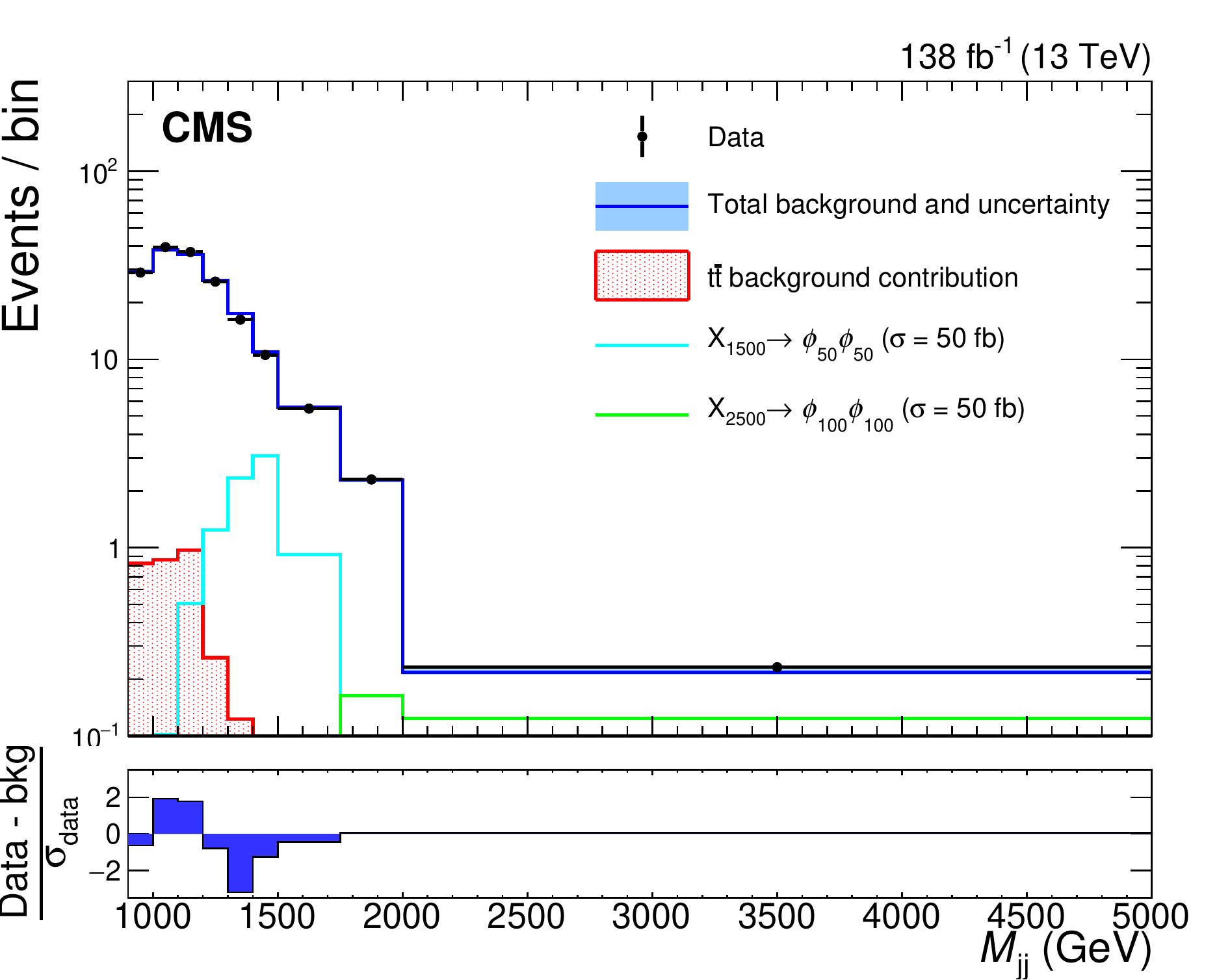}
    \caption{Distributions of average jet mass (left) and dijet mass (right), and background estimate of the combined search regions after the final fit is performed. The blue (solid) line represents the sum of the estimated QCD and \ttbar backgrounds, and the red filled histogram shows the \ttbar contribution alone. The shaded areas around the background estimate in the upper panels represent the total uncertainty in the total background estimate in that bin. The shapes of two representative signals, each normalized to cross sections of 50\unit{fb}, are indicated by solid colored lines. The lower panel shows the difference between the observed data and the background prediction, divided by $\sigma_{\text{data}}$, the statistical uncertainty of the data in each bin.}
    \label{EST1}
\end{figure*}

\begin{figure*}[hbt!]
    \centering
     \includegraphics[width=\linewidth]{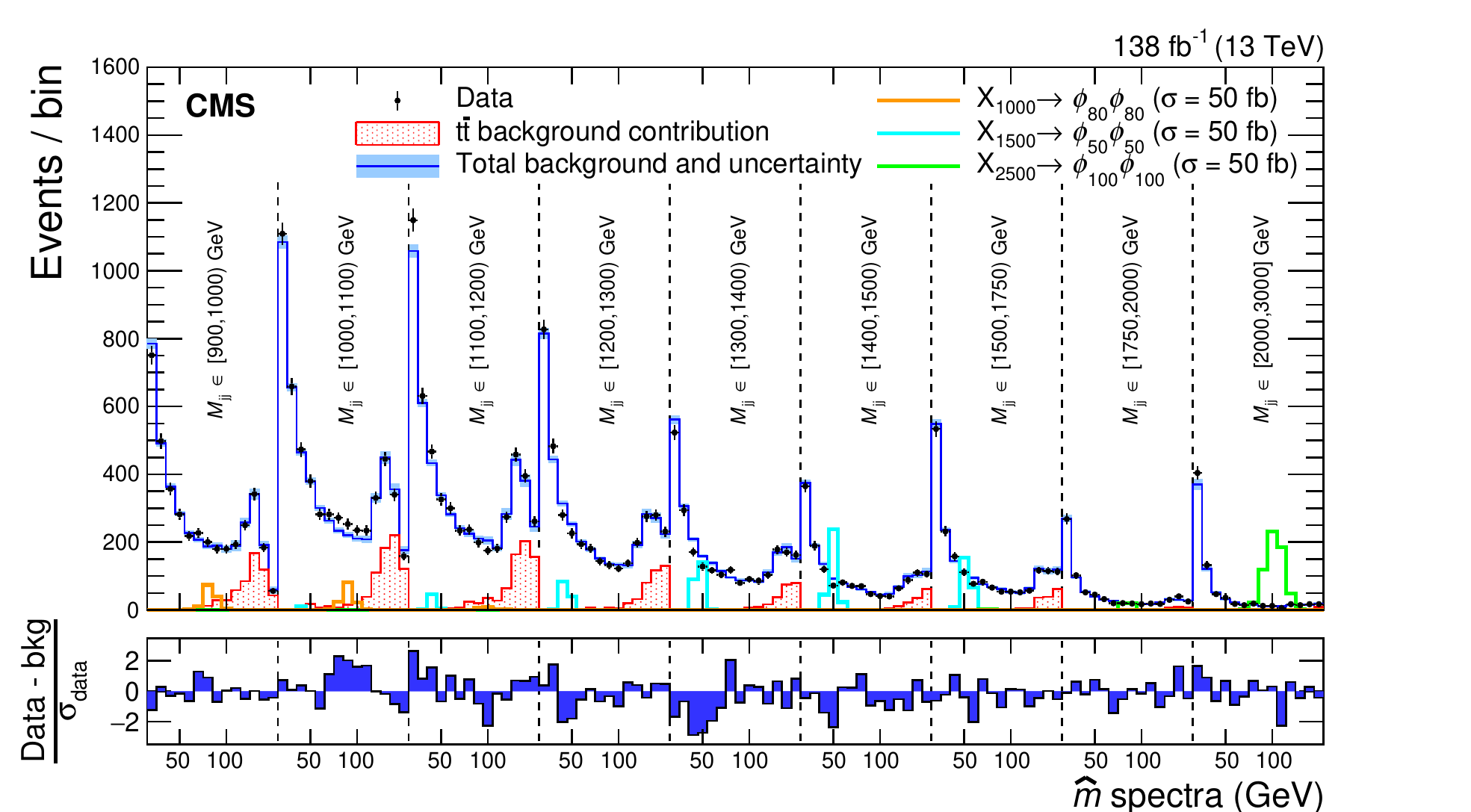}
    \caption{The average jet mass distributions in consecutive dijet mass intervals. The vertical dashed lines separate the average jet mass distributions in each bin of \Mjj. The individual bins within such subdivisions correspond to the $\hat{m}$ spectrum (from 15 to 200\GeV), as seen in Fig.~\ref{EST1} (\cmsLeft). Representative signal shapes are also shown; we note that they peak in the $\hat{m}$ spectrum within subdivisions, and may appear in multiple \Mjj bins. The blue (solid) line represents the sum of the estimated QCD and \ttbar backgrounds, and the red filled histogram shows the \ttbar contribution alone. The shaded areas around the background estimate in the upper panels represent the total uncertainty in the total background estimate in that bin. The shapes of three representative signals, each normalized to cross sections of 50\unit{fb} are indicated by solid colored lines. The lower panel shows the difference between the observed data and the background prediction, divided by $\sigma_{\text{data}}$, the statistical uncertainty of the data in each bin.}
    \label{EST2}
\end{figure*}

The final fits, for both the background-only and signal-plus-background hypotheses, simultaneously maximize a binned maximum likelihood over the TSR and LSR in all years.  When fitting for signal and background, the signal strength is the same in all regions. To validate the robustness of the fit, a goodness-of-fit test utilizing a binned likelihood ratio with respect to the saturated model as the test statistic is performed, yielding an overall goodness-of-fit with a p-value of 0.3. Bias tests are also performed, where the bias tests use data simulated from the background estimate with a variety of simulated signals injected (including the null hypothesis). The simulated data used both the arctangent representation of $F(\mjtwo)$ and equivalent third-order polynomials. The mean of the distribution of signal strengths obtained from a large number of trials is found to deviate by less than 0.5 standard deviations from the injected signal strength, confirming the absence of any significant bias. The results of the fit for the combined search regions are shown in Fig.~\ref{EST1} as projections onto the individual $\hat{m}$ and \Mjj distributions, and in Fig.~\ref{EST2} as average jet mass distribution in consecutive dijet mass intervals, showing also the difference between the data and the background estimate. The largest excess corresponds to $m_\PGf$ and $m_\PX$ of about 75\GeV and 1\TeV, respectively, with a local significance of 3.1 standard deviations and a global significance of 1.3 standard deviations. Here the global significance takes into account the look-elsewhere effect~\cite{GrossVitells}, using pseudo-experiments to measure the probability that the background hypothesis produces a signal-like effect with at least the observed local significance, anywhere within the sensitive range of $m_\PGf$ and $m_\PX$.

\section{Results}
\begin{figure*}[ht!p]
    \centering
    \includegraphics[width=0.95\linewidth]{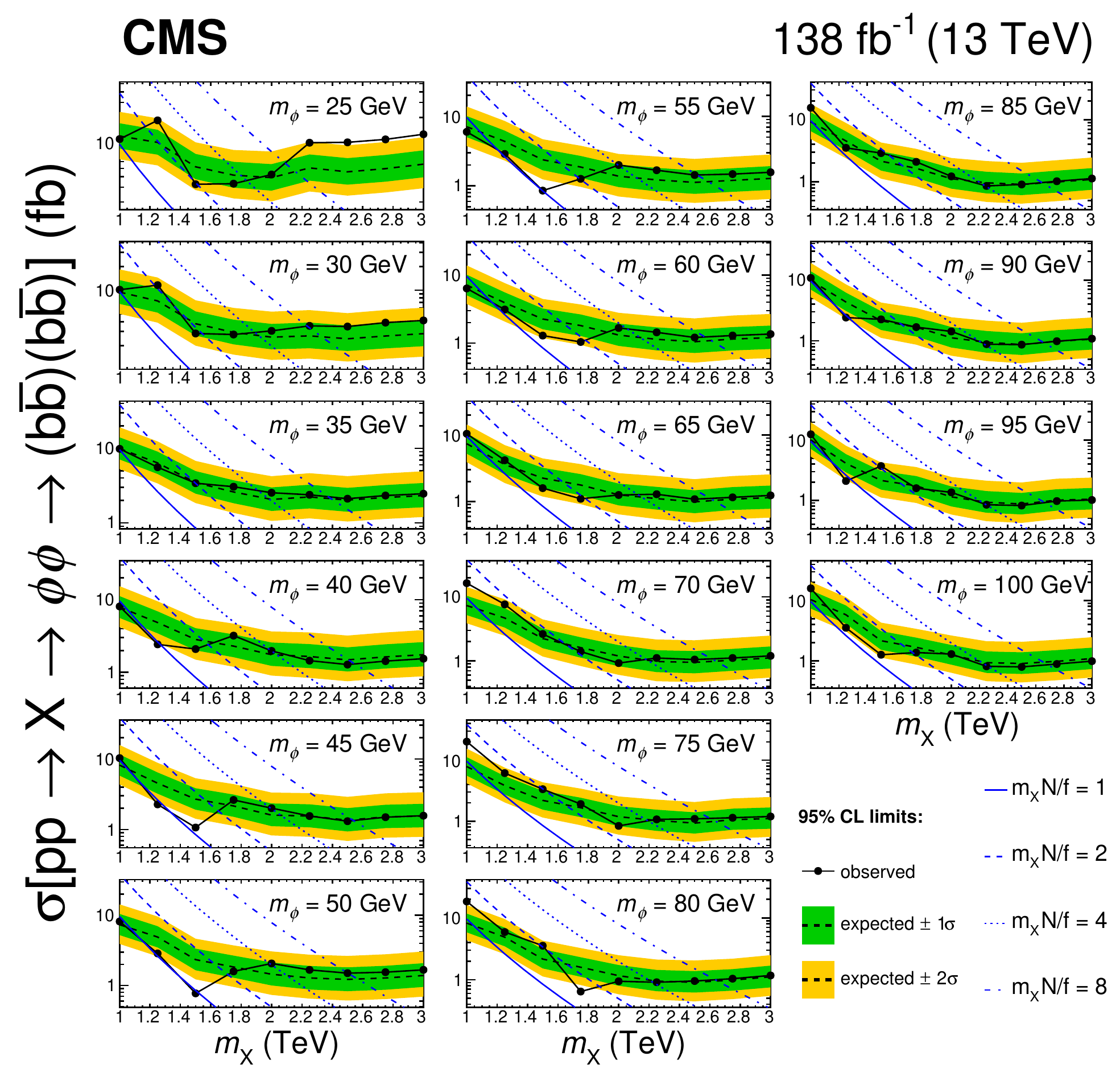}
    \caption{Upper limits at 95\% \CL on the cross section of the process $\Pp\Pp \to \PX \to \PGf\PGf \to (\PQb\PAQb)(\PQb\PAQb)$, as a function of the mass of $m_\PX$, for different values of $m_\PGf$. Both the $\PX \to \PGf\PGf$ and $\PGf \to \PQb \PAQb$ branching fractions are assumed to be 100\%. Each subpanel shows the limits for a fixed value of $m_\PGf$. The observed limits are shown as solid black lines with markers, while the expected limits are dotted. The yellow (outer) and green (inner) bands represent one and two standard deviation intervals. The theoretical cross section for different values of the parameter $m_{\PX} N/f$ are shown with dotted and dashed curves.}
    \label{LIM}
\end{figure*}

The results of the fit are used to set 95\% confidence level (\CL) upper limits on $\sigma(\Pp \Pp \to \PX)$, assuming a 100\% branching fraction for $\PX \to \PGf\PGf \to (\PQb\PAQb)(\PQb\PAQb)$. Upper limits are computed under a modified frequentist approach, using the \CLs criterion~\cite{CLS2,CLS1} with the profile likelihood ratio used as the test statistic with the asymptotic approximation~\cite{Cowan:2010js}. Observed limits are shown as a function of $m_\PGf$ and $m_\PX$, and compared to the theoretical estimates of $\sigma(\PX \to\PGf\PGf)$ for a set of $(m_{\PX} N)/f$ values in Fig.~\ref{LIM}. The upper limits on the process $\PX \to \PGf\PGf \to (\PQb\PAQb)(\PQb\PAQb)$ process range from 30 to 1\unit{fb}, depending on $m_\PGf$ and $m_\PX$.

\ifthenelse{\boolean{cms@external}}{}{\clearpage}
\section{Summary}
A search for massive resonances (\PX) decaying to pairs of spin-0 bosons (\PGf) that themselves decay into \PQb quark-antiquark pairs has been presented.  The analysis is restricted to the case where the mass ratio of the resonance and the scalar bosons is such that each pair of \PQb quarks is reconstructed as a single large-radius jet.  Data from proton-proton collisions at the LHC at $\sqrt{s}=13\TeV$ collected in 2016--2018 with the CMS detector, corresponding to an integrated luminosity of 138\fbinv, have been used. Upper limits are set at 95\% confidence level on the product of production cross section and branching fraction as a function of mass for $\PX \to \PGf\PGf\to (\PQb\PAQb)(\PQb\PAQb)$, where both the $\PX \to \PGf\PGf$ and $\PGf \to \PQb \PAQb$ branching fractions are assumed to be 100\%.
These are the first limits on this process, and range between 30 and 1\unit{fb} for a \PGf mass in the range 25--100\GeV and an \PX mass in the range 1--3\TeV.

\begin{acknowledgments}
    We congratulate our colleagues in the CERN accelerator departments for the excellent performance of the LHC and thank the technical and administrative staffs at CERN and at other CMS institutes for their contributions to the success of the CMS effort. In addition, we gratefully acknowledge the computing centers and personnel of the Worldwide LHC Computing Grid and other centers for delivering so effectively the computing infrastructure essential to our analyses. Finally, we acknowledge the enduring support for the construction and operation of the LHC, the CMS detector, and the supporting computing infrastructure provided by the following funding agencies: BMBWF and FWF (Austria); FNRS and FWO (Belgium); CNPq, CAPES, FAPERJ, FAPERGS, and FAPESP (Brazil); MES and BNSF (Bulgaria); CERN; CAS, MoST, and NSFC (China); MINCIENCIAS (Colombia); MSES and CSF (Croatia); RIF (Cyprus); SENESCYT (Ecuador); MoER, ERC PUT and ERDF (Estonia); Academy of Finland, MEC, and HIP (Finland); CEA and CNRS/IN2P3 (France); BMBF, DFG, and HGF (Germany); GSRI (Greece); NKFIA (Hungary); DAE and DST (India); IPM (Iran); SFI (Ireland); INFN (Italy); MSIP and NRF (Republic of Korea); MES (Latvia); LAS (Lithuania); MOE and UM (Malaysia); BUAP, CINVESTAV, CONACYT, LNS, SEP, and UASLP-FAI (Mexico); MOS (Montenegro); MBIE (New Zealand); PAEC (Pakistan); MSHE and NSC (Poland); FCT (Portugal); JINR (Dubna); MON, RosAtom, RAS, RFBR, and NRC KI (Russia); MESTD (Serbia); MCIN/AEI and PCTI (Spain); MOSTR (Sri Lanka); Swiss Funding Agencies (Switzerland); MST (Taipei); ThEPCenter, IPST, STAR, and NSTDA (Thailand); TUBITAK and TAEK (Turkey); NASU (Ukraine); STFC (United Kingdom); DOE and NSF (USA).
    
    \hyphenation{Rachada-pisek} Individuals have received support from the Marie-Curie program and the European Research Council and Horizon 2020 Grant, contract Nos.\ 675440, 724704, 752730, 758316, 765710, 824093, 884104, and COST Action CA16108 (European Union); the Leventis Foundation; the Alfred P.\ Sloan Foundation; the Alexander von Humboldt Foundation; the Belgian Federal Science Policy Office; the Fonds pour la Formation \`a la Recherche dans l'Industrie et dans l'Agriculture (FRIA-Belgium); the Agentschap voor Innovatie door Wetenschap en Technologie (IWT-Belgium); the F.R.S.-FNRS and FWO (Belgium) under the ``Excellence of Science -- EOS" -- be.h project n.\ 30820817; the Beijing Municipal Science \& Technology Commission, No. Z191100007219010; the Ministry of Education, Youth and Sports (MEYS) of the Czech Republic; the Deutsche Forschungsgemeinschaft (DFG), under Germany's Excellence Strategy -- EXC 2121 ``Quantum Universe" -- 390833306, and under project number 400140256 - GRK2497; the Lend\"ulet (``Momentum") Program and the J\'anos Bolyai Research Scholarship of the Hungarian Academy of Sciences, the New National Excellence Program \'UNKP, the NKFIA research grants 123842, 123959, 124845, 124850, 125105, 128713, 128786, and 129058 (Hungary); the Council of Science and Industrial Research, India; the Latvian Council of Science; the Ministry of Science and Higher Education and the National Science Center, contracts Opus 2014/15/B/ST2/03998 and 2015/19/B/ST2/02861 (Poland); the Funda\c{c}\~ao para a Ci\^encia e a Tecnologia, grant CEECIND/01334/2018 (Portugal); the National Priorities Research Program by Qatar National Research Fund; the Ministry of Science and Higher Education, projects no. 0723-2020-0041 and no. FSWW-2020-0008 (Russia); MCIN/AEI/10.13039/501100011033, ERDF ``a way of making Europe", and the Programa Estatal de Fomento de la Investigaci{\'o}n Cient{\'i}fica y T{\'e}cnica de Excelencia Mar\'{\i}a de Maeztu, grant MDM-2017-0765 and Programa Severo Ochoa del Principado de Asturias (Spain); the Stavros Niarchos Foundation (Greece); the Rachadapisek Sompot Fund for Postdoctoral Fellowship, Chulalongkorn University and the Chulalongkorn Academic into Its 2nd Century Project Advancement Project (Thailand); the Kavli Foundation; the Nvidia Corporation; the SuperMicro Corporation; the Welch Foundation, contract C-1845; and the Weston Havens Foundation (USA).
\end{acknowledgments}

\bibliography{auto_generated}
\cleardoublepage \appendix\section{The CMS Collaboration \label{app:collab}}\begin{sloppypar}\hyphenpenalty=5000\widowpenalty=500\clubpenalty=5000\cmsinstitute{Yerevan~Physics~Institute, Yerevan, Armenia}
A.~Tumasyan
\cmsinstitute{Institut~f\"{u}r~Hochenergiephysik, Vienna, Austria}
W.~Adam\cmsorcid{0000-0001-9099-4341}, J.W.~Andrejkovic, T.~Bergauer\cmsorcid{0000-0002-5786-0293}, S.~Chatterjee\cmsorcid{0000-0003-2660-0349}, K.~Damanakis, M.~Dragicevic\cmsorcid{0000-0003-1967-6783}, A.~Escalante~Del~Valle\cmsorcid{0000-0002-9702-6359}, R.~Fr\"{u}hwirth\cmsAuthorMark{1}, M.~Jeitler\cmsAuthorMark{1}\cmsorcid{0000-0002-5141-9560}, N.~Krammer, L.~Lechner\cmsorcid{0000-0002-3065-1141}, D.~Liko, I.~Mikulec, P.~Paulitsch, F.M.~Pitters, J.~Schieck\cmsAuthorMark{1}\cmsorcid{0000-0002-1058-8093}, R.~Sch\"{o}fbeck\cmsorcid{0000-0002-2332-8784}, D.~Schwarz, S.~Templ\cmsorcid{0000-0003-3137-5692}, W.~Waltenberger\cmsorcid{0000-0002-6215-7228}, C.-E.~Wulz\cmsAuthorMark{1}\cmsorcid{0000-0001-9226-5812}
% ...
\vskip\cmsinstskip
% ...
1:~Also at TU Wien, Wien, Austria\\
% ...
\end{sloppypar}
\end{document}